\documentclass[12pt]{article}
\usepackage{fancyhdr}

\usepackage{mathrsfs}
\usepackage[T1]{fontenc}
\usepackage{mathpazo}
\usepackage{setspace}
\usepackage{amsfonts}
\usepackage{amssymb}
\usepackage{amsmath}
\usepackage{epsfig}
\usepackage{latexsym}
\usepackage{color}
\usepackage{graphicx}
\usepackage{nicefrac}
\usepackage[latin1]{inputenc}
\usepackage{slashed}
\usepackage{multirow}
\usepackage{fancybox}
\usepackage{cite}
\usepackage{comment}
\usepackage{soul}
\usepackage{color}

\usepackage[all,cmtip]{xy}
\usepackage{hyperref}

\def\hybrid{\topmargin -20pt    \oddsidemargin 0pt
        \headheight 0pt \headsep 0pt
        \textwidth 6.25in       
        \textheight 9.25in       
        \marginparwidth .875in
        \parskip 5pt plus 1pt   \jot = 1.5ex}

\hybrid

\def\baselinestretch{1.2}

\catcode`\@=11

\def\marginnote#1{}
\newcount\hour
\newcount\minute
\newtoks\amorpm
\hour=\time\divide\hour by60
\minute=\time{\multiply\hour by60 \global\advance\minute by-\hour}
\edef\standardtime{{\ifnum\hour<12 \global\amorpm={am}%
        \else\global\amorpm={pm}\advance\hour by-12 \fi
        \ifnum\hour=0 \hour=12 \fi
        \number\hour:\ifnum\minute<10 0\fi\number\minute\the\amorpm}}
\edef\militarytime{\number\hour:\ifnum\minute<10 0\fi\number\minute}

\def\draftlabel#1{{\@bsphack\if@filesw {\let\thepage\relax
   \xdef\@gtempa{\write\@auxout{\string
      \newlabel{#1}{{\@currentlabel}{\thepage}}}}}\@gtempa
   \if@nobreak \ifvmode\nobreak\fi\fi\fi\@esphack}
        \gdef\@eqnlabel{#1}}
\def\@eqnlabel{}
\def\@vacuum{}
\def\draftmarginnote#1{\marginpar{\raggedright\scriptsize\tt#1}}

\def\draft{\oddsidemargin -.5truein
        \def\@oddfoot{\sl preliminary draft \hfil
        \rm\thepage\hfil\sl\today\quad\militarytime}
        \let\@evenfoot\@oddfoot \overfullrule 3pt
        \let\label=\draftlabel
        \let\marginnote=\draftmarginnote
   \def\@eqnnum{(\theequation)\rlap{\kern\marginparsep\tt\@eqnlabel}%
\global\let\@eqnlabel\@vacuum}  }

\def\preprint{\twocolumn\sloppy\flushbottom\parindent 2em
        \leftmargini 2em\leftmarginv .5em\leftmarginvi .5em
        \oddsidemargin -.5in    \evensidemargin -.5in
        \columnsep .4in \footheight 0pt
        \textwidth 10.in        \topmargin  -.4in
        \headheight 12pt \topskip .4in
        \textheight 6.9in \footskip 0pt
        \def\@oddhead{\thepage\hfil\addtocounter{page}{1}\thepage}
        \let\@evenhead\@oddhead \def\@oddfoot{} \def\@evenfoot{} }

\def\numberbysection{\@addtoreset{equation}{section}
        \def\theequation{\thesection.\arabic{equation}}}

\def\underline#1{\relax\ifmmode\@@underline#1\else
        $\@@underline{\hbox{#1}}$\relax\fi}

\def\titlepage{\@restonecolfalse\if@twocolumn\@restonecoltrue\onecolumn
     \else \newpage \fi \thispagestyle{empty}\c@page\z@
        \def\thefootnote{\fnsymbol{footnote}} }

\def\endtitlepage{\if@restonecol\twocolumn \else \newpage \fi
        \def\thefootnote{\arabic{footnote}}
        \setcounter{footnote}{0}}  

\catcode`@=12
\relax

\def\figcap{\section*{Figure Captions\markboth
        {FIGURECAPTIONS}{FIGURECAPTIONS}}\list
        {Figure \arabic{enumi}:\hfill}{\settowidth\labelwidth{Figure
999:}
        \leftmargin\labelwidth
        \advance\leftmargin\labelsep\usecounter{enumi}}}
 \relax
\def\tablecap{\section*{Table Captions\markboth
        {TABLECAPTIONS}{TABLECAPTIONS}}\list
        {Table \arabic{enumi}:\hfill}{\settowidth\labelwidth{Table
999:}
        \leftmargin\labelwidth
        \advance\leftmargin\labelsep\usecounter{enumi}}}
 \relax
\def\reflist{\section*{References\markboth
        {REFLIST}{REFLIST}}\list
        {[\arabic{enumi}]\hfill}{\settowidth\labelwidth{[999]}
        \leftmargin\labelwidth
        \advance\leftmargin\labelsep\usecounter{enumi}}}
 \relax
\makeatletter
\newcounter{pubctr}
\def\publist{\@ifnextchar[{\@publist}{\@@publist}}
\def\@publist[#1]{\list
        {[\arabic{pubctr}]\hfill}{\settowidth\labelwidth{[999]}
        \leftmargin\labelwidth
        \advance\leftmargin\labelsep
        \@nmbrlisttrue\def\@listctr{pubctr}
        \setcounter{pubctr}{#1}\addtocounter{pubctr}{-1}}}
\def\@@publist{\list
        {[\arabic{pubctr}]\hfill}{\settowidth\labelwidth{[999]}
        \leftmargin\labelwidth
        \advance\leftmargin\labelsep
        \@nmbrlisttrue\def\@listctr{pubctr}}}
 \relax
\makeatother
\newskip\humongous \humongous=0pt plus 1000pt minus 1000pt

\newif\ifdtup

\relax

\def\be{\begin{equation}}
\def\ee{\end{equation}}
\def\ba{\begin{eqnarray}}
\def\ea{\end{eqnarray}}

\def\del{\partial}

\def\k{\kappa}

\def\a{\alpha}
\def\A{\Alpha}
\def\b{\beta}
\def\B{\Beta}
\def\g{\gamma}
\def\G{\Gamma}
\def\d{\delta}
\def\D{\Delta}

\def\E{\Epsilon}

\def\C{\Chi}
\def\th{\theta}

\def\m{\mu}

\def\om{\omega}

\def\l{\lambda}

\def\s{\sigma}

\def\cN{{\cal N}}

\def\no{\noindent}

\def\qq{\qquad}

\def\IR{\relax{\rm I\kern-.18em R}}

\def \A { {\bar A} }

\def \ha {{1\over 2}}

\def \ov {\over}

\def\diag{{\rm diag}}
\def\const{{\rm const.}}

\def\IR{\relax{\rm I\kern-.18em R}}
\def\IL{\relax{\rm I\kern-.18em L}}

\def\inv{^{\raise.15ex\hbox{${\scriptscriptstyle -}$}\kern-.05em 1}}

\def\A{\mathcal{A}}
\def\B{\mathcal{B}}
\def\C{\mathcal{C}}
\def\D{\mathcal{D}}
\def\E{\mathcal{E}}
\def\F{\mathcal{F}}

\def\beq{\begin{equation}}
\def\eeq{\end{equation}}
\def\bea{\begin{eqnarray}}
\def\eea{\end{eqnarrat}}
\def\eqn{\eqref}
 
\begin{document}

\renewcommand{\theequation}{\thesection.\arabic{equation}}
\csname @addtoreset\endcsname{equation}{section}

 \def\baselinestretch{1.2}
 \noindent

\begin{titlepage}
\begin{center}


~

\vskip 1 cm

{\large \bf  Integrable $\lambda$-deformations: \\
 Squashing Coset CFTs and  $AdS_5\times S^5$ }

\vskip 0.4in

{\bf Saskia Demulder},$^1$ {\bf Konstantinos Sfetsos}$^2$~and~{\bf Daniel C. Thompson}$^1$\vskip 0.1in
{\em
{\it ${}^1$ Theoretische Natuurkunde, Vrije Universiteit Brussel,\\
\& The International Solvay Institutes,\\
Pleinlaan 2, B-1050 Brussels, Belgium.\\ 
\vskip
0.1in
${}^2$Department of Nuclear and Particle Physics\\
Faculty of Physics, University of Athens\\
Athens 15784, Greece\\
 {{\tt \footnotesize saskia.demulder@vub.ac.be}}, {\tt \footnotesize ksfetsos@phys.uoa.gr}}, {\tt \footnotesize daniel.thompson@vub.ac.be}}
\vskip .5in
\end{center}

\centerline{\bf Abstract}
\no
We examine integrable $\l$-deformations of $SO(n+1)/SO(n)$ coset CFTs 
 and their analytic continuations. We provide an interpretation of the deformation as a squashing of the corresponding coset $\sigma$-model's  target space.
 We  realise the $\l$-deformation for $n=5$ case as a solution to supergravity supported by non-vanishing five-form
 and  dilaton.  This interpolates between the coset CFT $SO(4,2)/SO(4,1)\times SO(6)/SO(5)$ constructed as a gauged WZW model and
 the non-Abelian T-dual of the $AdS_5\times S^5$ spacetime.

\no

\newpage

\tableofcontents

\noindent

\vskip .4in

\end{titlepage}
\vfill
\eject

\def\baselinestretch{1.2}
\baselineskip 20 pt
\noindent

\def\nn{{\tt{n}}}
\section{Introduction}

   $\cN=4$  supersymmetric Yang-Mills theory with gauge group  $SU(N_c)$  is in many ways an ideal theoretical
   laboratory to test our understanding of gauge theory and Quantum Field Theory;   it is complicated enough to capture many
   aspects of gauge theory dynamics yet rendered simple enough to be tractable.
   A particular limit, introduced by 't Hooft \cite{'tHooft:1973jz} is to consider the
   theory when the number of colours $N_c$ is taken very large with the coupling $ g_{YM}^2 N_c$ fixed.
   In this  limit only certain Feynman diagrams, those that can be drawn on a sphere in 't Hooft's double line notation,
   dominate and the theory becomes even simpler.  A remarkable property that has come to the fore since 2002 is that in
   this planar limit, the theory possesses a much enhanced underlying symmetry; in fact it is integrable.  This behaviour
   was first seen in \cite{Minahan:2002ve} where the difficult question of determining the spectrum of certain gauge invariant
   operators is mapped to an exactly solvable (integrable) system: the one-dimensional spin chain first introduced by Heisenberg in 1928 \cite{Heisenberg:1928mqa} and subsequently solved by Bethe \cite{Bethe:1931hc}.
   A second angle of attack on this gauge theory is provided by the AdS/CFT conjecture \cite{Maldacena:1997re} and   at strong coupling,
   integrability has  a very elegant manifestation in the holographic dual description; it corresponds to the existence of an infinite
   set of non-local conserved charges and   corresponding  underlying integrability of the two-dimensional non-linear
   $\s$-model describing strings in the $AdS_5 \times S^5$ spacetime \cite{Bena:2003wd}.

A very natural question we are led to ask is if we can relax some of the assumptions of supersymmetry whilst preserving
integrability?  One reason to hope this is possible is that it is known even QCD itself exhibits
some integrability  in certain limits \cite{Faddeev:1994zg,Lipatov:1994xy}. $\cN=4$ Yang-Mills theory admits a class of marginal deformations contained with those of \cite{Leigh:1995ep},
known as (real) $\beta$-deformations, that modify the superpotential, preserve only the minimal amount of (conformal) supersymmetry
but yet preserve integrability.
From the holographic perspective  the geometry describing these $\beta$-deformations
can be obtained by the application of $T$-dualities in TsT transformations \cite{Lunin:2005jy}.\footnote{One can consider including S-dualities and to find geometries corresponding to complex $\beta$-deformations but these are known to be non-integrable  \cite{Giataganas:2013dha}. } Strings propagating in this spacetime retain
the properties of integrability  \cite{Frolov:2005dj}.\footnote{  We can abolish all together the property of integrability but retain that of minimal
(conformal) supersymmetry. An example of that is $AdS_5 \times T_{1,1}$ \cite{Klebanov:1998hh},
where the propagation of strings can in fact be chaotic \cite{Basu:2011di}. However, in this case
the background is not obtained by a deformation of an integrable one.}
Can we go further in this direction and find deformations of the gauge theory
which still remain integrable but yet preserve no supersymmetry at all? If so, what is the corresponding holographic dual geometry?

 The past two years have seen the development of two new and related classes of integrable deformations  which we shall
 refer to as  $\eta$- and $\lambda$-deformations. The
 $\eta$-deformations were introduced by Delduc, Magro and Vicedo in \cite{Delduc:2013fga,Delduc:2013qra} and are based on Yang--Baxter
 $\sigma$-models proposed by Klimcik \cite{Klimcik:2002zj}.
These are described by $\sigma$-models propagating in a spacetime that is deformed from the original $AdS_5 \times S^5$
in a way that preserves no supersymmetries and  only an Abelian isometry group.
Despite this reduced symmetry, this deformation does preserve classical integrability of the $\sigma$-model.

The second class of  $\lambda$-deformations were proposed in \cite{Sfetsos:2013wia}  and are obtained by a gauging
procedure applied to a combination of a principal chiral model (PCM) and a Wess-Zumino-Witten (WZW) model \cite{Witten:1983ar} for a
semi-simple compact group.
After integrating out all non-propagating gauge fields one finds some resulting $\sigma$-models that have a
seemingly very complicated target spaces but remarkably also admit a Lax pair formulation and an underlying Poisson
(Dirac) bracket algebra of two commuting classical Kac-Moody algebras \cite{Sfetsos:2013wia} and corresponding Yangian \cite{Itsios:2014vfa}. The deformation parameter is given
in terms of the radius $\kappa$ of the PCM and the level $k$ of the WZW by
\be
\lambda = \frac{k}{k+ \k^2} \ .
\ee
For small values of $\lambda$ this deformation looks like a current-current deformation of the WZW model whilst in the
limit $\lambda \rightarrow 1$ then one arrives at the PCM but recast in a certain set of (non-Abelian-)T-dual variables.

Moreover, it has been shown that by utilizing a limiting procedure one constructs
models interpolating between gauged WZW models 
\cite{Bardakci:1987ee,Gawedzki:1988hq,Karabali:1988au} and the (non-Abelian) T-duals of the PCM on geometric coset spaces \cite{Sfetsos:2013wia}. The Lax pair formulation for $\l$-deformations based on symmetric spaces has been explicitly given in  \cite{Hollowood:2014rla}. 
In this case, the deformation is driven by parafermion bilinears which are the natural object with definite chirality
on shell.
Deformations of specific coset cases will be considered in detail in this paper and in particular we will focus on the
coset $SO(\nn+1)/SO(\nn)$ and its analytic continuations.

 Whilst some aspects of the relation between the  $\eta$- and $\lambda$-deformations are not completely understood,
 at the level of Poisson algebras   they correspond to
complementary ranges for a deformation parameter of the same algebra  \cite{Balog:1993es,Rajeev:1988hq}, see the discussion in
\cite{Delduc:2014uaa}.  At the quantum level it is expected that both types of deformations  can be understood in terms of quantum group deformations
 of the S-matrix symmetries. A sequence of works \cite{Hoare:2011nd,Hoare:2011wr,Hoare:2012fc,Hoare:2013ysa} have studied
 how the symmetries of the world-sheet S-matrix  may be deformed to a quantum group  whilst still satisfying S-matrix axioms.
 The deformation is labeled by a parameter $q$ and there are two cases to consider.  First is $q = e^{\eta} \in \mathbb{R}$
 which is thought to correspond to the $\eta$-deformation -- this has been supported by a matching in the large tension limit of the tree-level bosonic S-matrix for world-sheet scattering on the $\eta$-deformed $\s$-model with the prediction from the  $q$-deformed S-matrix  \cite{Arutyunov:2013ega}.   
 The second case is when $q$ is a root of unity and it was conjectured
 in \cite{Hollowood:2014rla,Hollowood:2014qma} that the $\lambda$-deformed theories  give a world-sheet
 realization for this scenario with the  quantum group parameter related to the level of the WZW
 by $q = e^{\frac{i \pi}{k}}$.

One should sound a note of caution here.  Despite their natural constructions, it is not automatic that either the $\eta$- or the $\lambda$-deformations are marginal.   Indeed, within bosonic string theory these deformation are not marginal. The running of $\lambda$ was calculated for
a general group in \cite{Itsios:2014lca,Sfetsos:2014jfa} and found to be in agreement with CFT results (all loop in $\l$ and leading order
in $1/k$).
Based on that, and symmetry arguments, it has been argued that this $\s$-model action is the all loop effective action for the
non-Abelian bosonized Thirring model. For the Yang--Baxter $\sigma$-model, on which the $\eta$-deformations were based, the running of
the coupling was computed in \cite{Squellari:2014jfa}.

The question is if the inclusion of the fermionic content of the type-II superstring can render these deformations exactly marginal. At the level of the string $\sigma$-model
 one should check that the $\beta$-functions vanish i.e. that the supergravity equations for the background fields are solved. 
   Whilst it is relatively clear how to extract the NS sector
 of the target space geometry for these deformations, a much harder task is to extract the background RR-fields.
 This is particularly tricky to do from first principles since the constructions of  \cite{Delduc:2013fga} and
 \cite{Hollowood:2014qma} are based on super-coset formulations -- these are appropriate for exposing algebraic
 properties but do obscure the geometry.   At the moment the best route one has is to consider the NS-sector and
 try to bootstrap the solution for the RR-fields by directly solving the supergravity equations.  If such
 a supergravity embedding exists one can take it as  strong evidence for the consistency of these deformations.

For $\eta$-deformations the full supergravity embeddings were constructed for examples based on $AdS_2$ and $AdS_3$
in \cite{Lunin:2014tsa}. However, the complete supergravity embedding for $AdS_5\times S^5$ has not, at present, been established.
In this note we will focus on the $\lambda$-deformations for which some low-dimensional examples were recently given
supergravity embeddings \cite{Sfetsos:2014cea}. The primary purpose of this paper is to provide the full supergravity
embedding for the case of $\lambda$-deformations of $AdS_5\times S^5$.  Along the way we shall take the opportunity to
clarify a number of geometric properties of $\lambda$-deformed $\sigma$-models.  We shall show that the $\lambda$ deformations
of theories based on $SO(\nn+1)/SO(\nn)$ have a remarkably simple interpretation; they correspond to squashing of certain directions
in tangent space.  Whilst it is known that geometric cosets admit integrable squashings \cite{Orlando:2010yh,Kawaguchi:2011mz,Kawaguchi:2011pf,Orlando:2012hu}, here we are performing squashings on
conformal cosets (i.e. on gauged WZW models). After analytic continuation one finds the similar results for $AdS$ conformal
cosets.  Finally we are, with a simple Ansatz motivated by the underlying group theory, able to deduce the forms of the RR-fields.

The structure of this paper is as follows:  In Sec.~\ref{recap} we review and clarify some aspects of the $\lambda$-deformations
in general before specialising in Sec.~\ref{sphere} to the case of  $SO(\nn+1)/SO(\nn)$.  In  Sec.~\ref{wick} we briefly expound
on the continuations that give rise to deformations of $AdS$ conformal cosets and finally in Sec.~\ref{sugra} we present the full
supergravity embedding for the case of the $\lambda$-deformed $AdS_5\times S^5$.  For completeness we provide an appendix collating general results for the lower dimensional  cases of   $\nn=2,3$ and $4$.

\section{$\lambda$-deformations overview}
\label{recap}

We now review, and further develop, the results of \cite{Sfetsos:2013wia} relevant for this paper.  We begin with  some conventions.
We consider a general (bosonic) compact group $G$ and a corresponding group element
$g$ parametrized by $X^\m$, $\m=1,2,\dots , \dim(G)$. The right and  left invariant Maurer--Cartan forms, as well as the orthogonal
matrix (or adjoint action) relating them, are defined as
\be
\begin{aligned}
& L^A_\pm = -i\, {\rm Tr}(T^A g^{-1} \del_\pm g ) \ , \quad R^A_\pm =  -i\, {\rm Tr}(T^A \del_\pm g g^{-1})\ ,
\\
 & R^A_\m = D_{AB}L^B_\m\ ,  \qq D_{AB}={\rm Tr}(T_A g T_B g^{-1})\ .
 \label{jjd}
 \end{aligned}
\ee
The matrices $T^A$ obey $[T_A,T_B]=i f_{ABC} T_C$ and are normalized as ${\rm Tr}(T_A T_B)=\d_{AB}$.   In a coset $G/H$ we will let $T^a$, $a=1\dots \dim(H)$, be
generators for the subgroup and $T^\alpha$ , $\alpha=1\dots  \dim(G/H)$,  the remaining coset generators.
World-sheet light cone coordinates are $\sigma^\pm = \tau \pm \sigma$.

\subsection{$\lambda$-deformations for groups}
Although our main interest is in $\lambda$-deformations for certain symmetric spaces let us
begin by reviewing the simpler case of $\lambda$-deformations for groups since it will prove to be quite instructive.
Here the two ingredients are a bosonic principal chiral model (PCM) on the group manifold for an element $\tilde{g}\in G$,
\be
\label{eq:PCM}
S_{{\rm PCM}}(\tilde{g}) = \frac{\k^{2}}{\pi} \int_{\Sigma} \d_{AB} L_{+}^{A}(\tilde{g})  L_{-}^{B}(\tilde{g}) \  ,
\ee
and a  Wess-Zumino-Witten model (WZW)  for a second group element $g  \in G$  defined by
\be
\label{eq:WZW}
S_{{\rm WZW},k}(g) =
\frac{k}{2\pi} \int_{\Sigma} \delta_{AB}L^A_+(g) L^B_-(g) \,
  +\frac{k}{12\pi}\int_{{\cal B}} f_{ABC} L^A\wedge
L^B\wedge L^C\ ,
\ee
where ${\cal B}$ is an extension such that $\partial {\cal B} = \Sigma$.  To obtain the $\lambda$-deformation
one simply takes the sum of the PCM and the WZW  and applies a gauging procedure that eliminates half of the degrees of freedom.
More precisely one gauges the left action of $G_L$ in the PCM and the diagonal action of $G$ in the WZW with a single common
$G$-algebra valued one-form gauge field $A= i A^A T^A$.  This is done by minimal coupling in the PCM and by replacing the WZW with
the $G/G$ gauged-WZW model.  One can now choose a gauge fixing choice $\tilde{g}=\mathbb{1}$ such that the action becomes
\be
 S_{\rm tot} = S_{WZW}(g) + \frac{k}{\pi} \int A_+L_-^A - A_-^A R_+^A + A_+^AM_{AB} A_-^B \ ,
 \ee
 where the quadratic term in the gauge field couples to
 \be
 M_{AB}= \lambda^{-1} \delta_{AB} - D_{B A}(g)\ .
 \ee
The deformation parameter is given in terms of the radius $\kappa$ of the PCM and the level $k$ of the WZW by
\be
\lambda = \frac{k}{k+ \k^2} \ .
\ee
 To give this theory a non-linear $\sigma$-model interpretation we can continue by integrating out the gauge fields to give
  \be
 \label{eq:gaugeeqm}
 M A_- = - L_-  \ , \qquad M^T A_+ = R_+  \ .
 \ee
  Upon substitution of these equations the action becomes
  \be
   \label{eq:Sonshell}
  \begin{aligned}
 S_{\rm tot} &=  S_{WZW}(g) + \frac{k}{ \pi} \int  R_+^TM^{-1} L_- \ .
\\
  \end{aligned}
 \ee
 One can now go ahead and read off from this action the target space metric and the NS two-form potential.
Letting $M D = N$ we can see that the target space metric is
\be
\label{eq:metricgroup}
  \begin{aligned}
ds^2 &= k L^A (  \mathbb{1} +  D^T M^{-1} +  M^{-T}  D)_{AB} L^B   \\
&= k L^T  N^{-1} ( N N^T + N^{T} + N) N^{-T} L  \\
&= k (\lambda^{-2}- 1)  L^T  N^{-1} N^{-T} L  \\
& = k (\lambda^{-2}- 1) {\cal A}_{+}^T {\cal A}_{+}\ ,
 \end{aligned}
\ee
where ${\cal A}_{+}  $ is the push forward  i.e. $A_+ = {\cal A}_{+i} \partial_+ x^i$.
This shows that the push-forward of the on-shell values of the gauge fields define frame fields for the deformed sigma model in the case of groups.
An analogous calculation shows that the push forwards  ${\cal A}_{-}  $ also serve as frame fields.  Indeed from the equations of motion we have
\be
 {\cal A}_{-} = - M^{-1}D^T M^T  {\cal A}_{+} = \Lambda^T {\cal A}_+  \ , \quad \Lambda^T =  (1- \lambda D^T)^{-1}(\lambda - D^T)\ ,
\ee
 which defines a local frame rotation between these frame fields.

Just as is the case with gauged-WZW models, integrating out the gauge fields in a path integral produces a determinant factor giving rise to a non-trivial dilaton as
\be
\Phi = -\ha \ln \det M\ .
\ee

Whilst such a construction may seem esoteric it has quite a natural motivation;
it corresponds to performing a Buscher dualisation of the $G_L$ action in the PCM but upgrading the Lagrange multiplier term to a
fully dynamic sector - the WZW  \cite{Sfetsos:1994vz,Sfetsos:2013wia}.
Indeed one can consider a limit in which $k\gg 1$ and $\l\to \mathbb{1}$ in which we expand the group element of the WZW as
  \be
   g = \mathbb{1} + i { v_A T^A \ov k} + {\cal O}\left( 1\ov k^2 \right) \ .
 \label{glim}
  \ee
 Then the gauged WZW term reduces exactly to a Lagrange multiplier enforcing the gauge fields to be pure gauge as in the Buscher procedure.
In this limit the final action \eqn{eq:Sonshell} corresponds precisely to the non-Abelian T-dual of the PCM with respect to the $G_L$ isometry.

\subsection{$\lambda$-deformations for cosets}
Let us now consider the PCM on a geometric coset $G/H$.  This can be obtained either by a limiting procedure or a
gauging procedure but can be expressed by restricting the sum over generators in the PCM to a sum over coset generators
\be
\label{eq:PCMGmodH}
S_{{\rm PCM, G/H}}(\tilde{g}) = \frac{\k^{2}}{\pi} \int_{\Sigma} \d_{\a\b} L_{+}^{\a}(\tilde{g})  L_{-}^{\b}(\tilde{g}) \ .
\ee
This action develops a local $H$ invariance that can be fixed with an appropriate choice of   $\tilde{g}$.
We now repeat exactly the same steps as for the group; we supplement with a WZW model for an element $g\in G$ and gauge
the $G_L$ of the PCM and a $G_{\rm diag}$ of the WZW with a common gauge field.  Once again we fix $\tilde{g}=\mathbb{I}$ such that the total action is
\be
\label{eq:Stotcoset}
 S_{Tot} = S_{WZW}(g) + \frac{k}{\pi} \int A_+L_-^A - A_-^A R_+^A + A_+^AM_{AB} A_-^B \ ,
 \ee
 but now the quadratic matrix $M_{AB}$ distinguishes between subgroup directions\footnote{Roman lower case indices
correspond to subgroup and Greek to coset directions.}:
 \be
 \label{eq:Mcoset}
M_{AB} = E_{AB} - D_{BA} \ , \quad E_{AB} =  \left(\begin{array}{cc}  \mathbb{1}_{ab} & 0 \\ 0 &   \lambda^{-1} \mathbb{1}_{\a\b} \end{array}  \right) \ .
\ee
However, in gauge fixing $\tilde{g}=\mathbb{1}$ we have still not used up all the gauge freedom,
and it is clear from the form of eqs.~\eqref{eq:Stotcoset} and \eqref{eq:Mcoset} that the
final action retains a residual $H$ gauge symmetry that will require gauge fixing on $g$.

We can continue exactly as with the case of groups and integrate out the gauge fields  to yield again
  \be
   \label{eq:Sonshell coset}
  \begin{aligned}
 S_{Tot} &=  S_{WZW}(g) + \frac{k}{ \pi} \int  R^TM^{-1} L_- \ .
\\
  \end{aligned}
 \ee

To  read off the target space metric we need some care; a key  observation is that if we denote $N= MD$ and $\tilde{N} =D M$,
by virtue of the orthogonality of the adjoint action $D^TD = D D^T = \mathbb{I}$ we have the identity
\be
\label{eq:NNtildeidentity}
 N N^T + N^{T} + N  =    \tilde{N}^T \tilde{N} + \tilde{N}^{T} + \tilde{N}
= \left(\begin{array}{cc}  0 & 0 \\ 0 &  ( \lambda^{-2} -1)  \mathbb{1}_{\a\b} \end{array}  \right)  \ .
\ee
This can be used in a similar way to the derivation of eq.~\eqref{eq:metricgroup} to
show that the target space metric for  the $\sigma$-model in eq.~ \eqref{eq:Sonshell coset} is
    \be
    \label{eq:metriccoset}
   \begin{aligned}
 ds^2 &= k L^A (  \mathbb{1} +  D^T M^{-1} +  M^{-T}  D)_{AB} L^B   \\
   &= k  ( \lambda^{-2} -1)  (L^TN^{-1})^\a  \delta_{\a\b} (N^{-T} L)^\b\\
      &= k  ( \lambda^{-2} -1)  {\cal A}_+^\a  \delta_{\a\b}   {\cal A}_+^\b \ .
  \end{aligned}
 \ee
 One sees that  the push forward  to target space of the gauge field  ${\cal A}_{+}^\a  $ with {\em legs
in the coset directions} of the algebra defines a frame field.   An analogous calculation, making use again of eq.~\eqref{eq:NNtildeidentity},
establishes also  ${\cal A}_{-}^\a  $ to be a second set of frame fields. We must thus have
   \be\label{eq:Lambdadef}
     {\cal A}_{-}^{\a}=(\Lambda^{T} )_{\a\b} {\cal A}_{+}^{\b} \ , \quad (\Lambda^{T})_{\a\b} \Lambda_{\b\gamma} = \delta_{\a\g} \ .
       \ee
By comparison to the gauge field equations of motion we know that $\Lambda_{\a\b}$ must be no more than the
projection into coset indices of $\Lambda_{AB}=( -M D M^{-T})_{AB}$.

Our next goal is to describe exactly the form that these frame fields take, i.e. to examine the structure of the on-shell value of the gauge fields.
This can be achieved  by explicitly decomposing into subgroup and coset.
We define
\be
\label{eqref:adjdecomp}
D=
  \left(\begin{array}{cc}  d_{1} & d_{2} \\ d_{3} & d_{4} \end{array}  \right)  \ ,
\quad M =   \left(\begin{array}{c|c}  \mathbb{1}-  d^{T}_{1} & -d_{3}^{T}  \\ \hline -d_{2}^{T} & \l^{-1} \mathbb{1}- d_{4}^{T} \end{array}  \right)
\equiv    \left(\begin{array}{cc} {\bf A} &  {\bf B} \\ {\bf C} & {\bf D}\end{array} \right)
   \ee
in which the top left hand square block has dimension $\dim H$ and the bottom right $\dim(G/H)$. Note  that the blocks $d_i$, $i=1,2,3,4,$
are not   independent since the matrix $D$ has to be orthogonal.
The standard matrix inversion formula gives
 \be
 M^{-1}
 =  \left(\begin{array}{cc} {\bf Q}^{-1} &  - {\bf A}^{-1} {\bf B} {\bf P}^{{-1}} \\  - {\bf P}^{-1}{\bf C} {\bf A}^{-1}& {\bf P}^{-1}\end{array} \right)
 \ee
  where
  $ {\bf P} = {\bf D} - {\bf C} {\bf A}^{-1} {\bf B}$ and ${\bf Q} = {\bf A}- {\bf B}{\bf D}^{-1}{\bf C} $.
  The matrix  ${\bf P}$ will be very important and explicitly has elements
 \be
 \label{eq:Pmatrix}
 {\bf P}_{\a \b} = (\l^{-1}\mathbb{1} - d_{4}^{T})_{\a\b} -( d_{2}^{T}(1- d_{1}^{T})^{-1} d_{3}^{T})_{\a\b} \ .
 \ee 
 Using the relations between the blocks $d_i$, it can be shown that the following identity is obeyed
 \be
 ( {\bf P}-\l^{-1} \mathbb{1})  ( {\bf P}^T-\l^{-1} \mathbb{1})
 = ( {\bf P}^T-\l^{-1} \mathbb{1})  ( {\bf P}-\l^{-1} \mathbb{1})
 =  \mathbb{1}\ ,
 \label{PPT}
 \ee
 which also implies that $[ {\bf P}, {\bf P}^T]=0$. The proof of the second equality follows simply also from the
 fact that $ {\bf P}(g^{-1})= {\bf P}^T(g)$.

Expanding out the gauge field equations in this way gives the following expression for the frame fields ${\cal A}_{\pm}^\a$ entering into  eq.~\eqref{eq:Lambdadef}
 \be
 \begin{aligned}
   {\cal A}_{-} ^\a &=   -({\bf P}^{-1})^{\a\b} \left( L^{\b} - ({\bf CA}^{-1})^{\b a}L^{a} \right) \ ,
   \\
   {\cal A}_{+}^\a &=  ({\bf P}^{-T})^{\a\b}   \left( (\l^{-1}\mathbb{1} - {\bf P}^{T} )^{\b\g}  L^{\g} - ({\bf B}^{T}{\bf A}^{-T})^{\b a}L^{a}  \right)\ .
 \end{aligned}
 \ee 
 Consider for a moment the ${\cal A}_{-}$ frame fields; notice that the term in parenthesis consists of only the Maurer-Cartan forms
 and the matrices  ${\bf C}$ and ${\bf A}$ none of which by definition depends on the deformation parameter $\lambda$.
 Hence, the effect of the $\lambda$-deformation is entirely contained in the matrix ${\bf P}^{-1} $ acting as an overall dressing factor.
 
We can also calculate the matrix relating left and right frame fields in eq.~\eqref{eq:Lambdadef} and we see again the matrix ${\bf P}$ plays a crucial role
\be\label{eq:Lambda2}
(\Lambda^{T} )_{\a\b} = \left[  \lambda^{-1} \mathbb{1} + (1-\l^{-2}) {\bf P}^{-1}\right]_{\a\b} \ .
\ee
In the above manipulations the identity \eqn{PPT} has been particularly useful.

Just as is the case of $\l$-deformations for groups, integrating out  the gauge fields in a path integral produces a determinant
factor giving rise to a non-trivial dilaton.  For the case at hand the relevant quantity to consider is $\det M$ which can be computed by writing
\be
M = \left(
      \begin{array}{cc}
        {\bf A} & 0 \\
        {\bf C} & \mathbb{1} \\
      \end{array}
    \right) \left(
                     \begin{array}{cc}
                       \mathbb{1} & {\bf A}^{-1} {\bf B} \\
                       {\bf 0} & {\bf P} \\
                     \end{array}
                   \right)\ .
\ee
Then
 \be
 \label{eq:DilDet}
 \det M = \det {\bf A} \det{\bf P}  \quad  \Longrightarrow\quad \Phi = - \frac{1}{2}\ln \det {\bf A} - \frac{1}{2}\log    \det{\bf P}   \ .
 \ee
 By definition ${\bf A}$ is independent of $\lambda$ and precisely gives the contribution to the dilaton for an undeformed gauged-WZW whereas the matrix ${\bf P}$ depends on $\lambda$ and gives an additive contribution to the dilaton.\footnote{We remark that in the case of $\eta$-deformations, the dilaton does not obey a simple factorization as it does here; this  probably indicates that $\eta$ deformations can not arise via a gauging procedure.}
It is evident that to proceed further we need to understand the structure of this matrix ${\bf P}$ which one might anticipate having a rather complex coordinate dependence.   In fact this is not the case for the $\lambda$-deformations of $SO(\nn+1)/SO(\nn)$ cosets; the matrix ${\bf P}$ turns out to be coordinate independent and has eigenvalues that are simply  $\lambda^{-1} \pm 1$.  
However since ${\bf P}$ is gauge dependent, to see this explicitly requires some effort and is made easier by a  judicious gauge fixing choice to which  we now turn.

\section{$\lambda$-deformations  and Squashed Conformal Spheres }
\label{sphere}

Gauged WZW models for   $SO(\nn+1)/SO(\nn)$, which we shall call conformal cosets $CS^\nn$
to distinguish them from the geometric coset $S^\nn  = SO(\nn+1)/SO(\nn)$, and their analytic continuations
have been studied for quite some time.  For the case of $\nn = 2$ these were considered
in \cite{Bardacki:1990wj,Rocek:1991vk,Witten:1991yr} and famously given an interpretation by Witten as a
CFT description of a black-hole target space \cite{Witten:1991yr}.   Higher dimensional generalizations for
the cases of $\nn =3,4$  were constructed in \cite{Bars:1991pt,Fradkin:1991ie} and
\cite{Bars:1991zt}, respectively.  A second reason for the importance of such models comes from the
understanding that string theory in $AdS_5\times S^5$ can be reduced by a clever gauge fixing to a gauged WZW for the
conformal coset $CAdS_4\times CS^4$ supplemented by a suitable potential.  The elimination of gauge degrees of freedom
in this way goes by the name Pohlmeyer reduction \cite{Pohlmeyer:1975nb} and the application to superstrings in $AdS_5\times S^5$
was described in \cite{Grigoriev:2007bu} building on the earlier suggestions of \cite{Bakas:1995bm,FernandezPousa:1996hi}.

The first step is to specify a gauge fixing choice for the group element of $SO(\nn +1)$
that uses up the $SO(\nn)$ gauge invariance.
 The $SO(\nn+1)$ generators $T_{mn}$, with $m,n=1,2,\dots, \nn+1$ satisfy the commutation relations
\be
[T_{mn},T_{k\ell}] =\d_{nk} T_{m\ell} - \d_{mk} T_{n\ell} - \d_{n\ell} T_{m k} + \d_{m\ell} T_{nk}\ .
\ee
In the fundamental representation the matrix elements are
\be
(T_{mn})_{ab} = \delta_{ma}\delta_{nb} - \delta_{mb}\delta_{na} \ , \quad m,n,a,b = 1,2,\dots, \nn+1\ .
\ee

Then we can parametrize the gauged fixed group element in generalised Euler angles, $\th_i, i=1,2,\dots ,\nn-1 $ and $\phi$,
as in \cite{Fradkin:1991ie,Grigoriev:2007bu} (and advocated for the current context in  \cite{Hollowood:2014qma}) by
\be\label{gfixed}
g = \left(\prod_{i=1}^{\nn-1} g_i(\th_i) \right) g_{\nn}(2 \phi)  \left(\prod_{j=1}^{\nn-1}
g_{\nn-j}(\th_{\nn-j}) \right)\ ,
\ee
where
\be
g_k(x) = \exp(x T_{k, k+1})\ ,
 \ee
 generates a rotation in the $(k,k+1)$-plane.  This gauge fixing has a couple of extremely useful properties.
 First, as one can verify, the adjoint matrix reduces to
\be\label{eq:Dsimp}
D_{M N}[g]  \equiv   D_{mn, pq}[g] =   g_{n p } g_{m q } -  g_{m p }g_{n q }  \ .
\ee
This shows that the adjoint action in this gauge fixing coincides with
the anti-symmetric representation in this fixing.   The second is that, as was emphasized in  \cite{Grigoriev:2007bu}
that there exists an automorphism of the algebra (and, by exponentiation, on the group) that acts on a matrix ${\cal M}$ 
\be
({\cal M})_{ab} \rightarrow  (-1)^{a+b} ({\cal M})_{ab}  \ ,
\ee
and which clearly preserves the trace of a matrix.
Acting on the gauge fixed element in eq.~\eqref{gfixed} $g$ is sent to $g^{-1}$.
As a consequence, the three-form $\omega = (g^{-1}dg )^{\wedge 3}$
is odd under this automorphism so  $\mathrm{Tr} \omega = 0$ and the Wess--Zumino term can not contribute to the action.
For similar reasons, detailed in the appendix B of   \cite{Grigoriev:2007bu},
no contribution to an anti-symmetric Kalb--Ramond field can arise from integrating out the gauge field
in the gauged WZW model. This is in agreement with the explicit constructions for $\nn=3,4$ in
\cite{Bars:1991pt,Fradkin:1991ie,Bars:1991zt}.
Though our theory is modified by the deformation,
the couplings to the gauge field are also simply linear in $\partial g$  and so this argument
is also explains why we have no $B$-field in
the case at hand.\footnote{We note that having vanishing $B$-field is different to the case of $\eta$-deformations.
However, one should remember that the connection between the $\lambda$- and $\eta$- deformations
appears to involve performing a non-abelian T-duality transformation in which $B$-field can be traded for geometry.
We thank Ben Hoare for discussions on this point.}

For calculational purposes it is also useful to consider a second parametrization
of the group element as in \cite{Bars:1991zt}, namely that in which
\be
g = H t \ ,
\ee
where
\be
H = \left(
      \begin{array}{cc}
        1 & 0 \\
        0 & h \\
      \end{array}
    \right)\ ,\qq h= (\mathbb{1}+A)(\mathbb{1}- A)^{-1} \ ,
\ee
and
\be
t = \left(
      \begin{array}{cc}
        b &( b+1)  V^t \\
        - (b+1) V & \mathbb{1} - (b+1) VV^t \\
      \end{array}
\right)\ , \qq  b  ={1-V^2 \ov 1+V^2}\  ,
\ee
where $V$ is an $\nn$-dimensional vector and $A$ is an antisymmetric $\nn\times \nn$ matrix. In this way $t\in SO(\nn+1)$ and $h \in SO(\nn)$.
Hence we write
\ba
g=   \left(
     \begin{array}{cc}
        b & (b+1)  V^t \\
        -(b+1) \tilde{V} &  h - (b+1) \tilde V V^t \\
      \end{array} \right)  \  ,
    \quad \tilde{V}_\a = h_{\a\b} V_\b\ , \quad \a,\b= 1,2,\dots, \nn  \ .
\ea
One must then choose a gauge fixing to reduce to $\dim G - \dim H$
degrees of freedom in one-to-one correspondence with the   $\dim G - \dim H$
$H$-gauge invariants that can be built from $A$ and $V$.
The correspondence to the generalised Euler angles introduced above can be easily deduced. The general expressions are,

\be
A_{i,i+1} = \frac{\sin \th_i}{\cos \th_{i-1}\cos \th_{i}} \ ,  \quad i =1,2,\dots, \nn-1\ ,
\ee
with  $\th_0=0$, all other entries not fixed by symmetry equal to zero and,
   \be
 v_i = (-1)^{i+\nn} \tan\phi \cos\th_{i-1} \sin\th_i \sin\th_{i+1}\cdots \sin\th_{n-1}\ ,\qq i= 1,2,\dots, \nn\ ,
 \ee
  such that $ b = \cos 2 \phi $.

In terms of these quantities we can express the components of the adjoint matrix eq.~\eqref{eqref:adjdecomp} as
  \be
  \begin{aligned}
(d_{1})_{\a\b, \g\d} &=  b^2 \left[(d_4^{-T})_{\a\g} (d_4^{-T})_{\b\d}-  (d_4^{-T})_{\b\g} (d_4^{-T})_{\a\d} \right]
\\
(d_{1}^{-T})_{\a\b
, \g\d} &=  b^{-2} \left[ ((d_4)_{\a\g} (d_4)_{\b\d}-  (d_4)_{\b\g} (d_4)_{\a\d} \right] \ ,
     & \\
 (d_{2})_{\a\b
,   \g}&= (b+1) \tilde{V}_{[\a} h_{\b] \g} \ , \qquad    (d_{3})_{  \g,\a\b}=(b+1) h_{\g [ \a} V_{\b]}  \ ,
  & \\
  (d_4)_{\a\b}  &  = b h_{\a\b} + (b+1) \tilde{V}_\a V_b \ , \\
  (d_4^{-1}) _{\a\b} &=  b^{-1} h_{\b\a} -  b^{-1} (b+1) V_\a  \tilde{V}_b  \  .
  \end{aligned}
  \ee
Using this gauge fixing parametrization one can directly compute the matrix ${\bf P}$ matrix defined in eq.~\eqref{eq:Pmatrix}
for $SO(\nn+1)/SO(\nn)$. Indeed, eventually this takes the simple form
\be
 \label{eq:Pmatrix2}
 {\bf P}_{\a \b}
 = \left(\lambda^{-1} + (-1)^{\nn+\a + 1} \right)\delta_{\a\b} \ .
\ee
Notice that all coordinate dependence cancels out and that in this basis ${\bf P}$ is already diagonal.
This relation was verified explicitly for the physically relevant cases of $\nn=3,4,5$ and one anticipates,
though we did not show it analytically, that it holds in general given the form of the adjoint action in eq.~\eqref{eq:Dsimp}.

\no
 As a consequence one can now see that the effect of the $\lambda$-deformation is to perform
a squashing of the conformal coset's tangent space by a diagonal matrix acting on the frames. To be precise
let's denote the frame from which one computes the deformed metric by $e^\a_{(\l)}$. We have that
\be
e^\a_{({\l})}  = S^{\a\b}(\l) e^\b_{(0)} \ ,
\label{fram1}
\ee
where the lower index in $e^\a_{(0)}$ indicates that this is the frame for the undeformed metric.
The matrix $S(\l)$ is diagonal and given explicitly by
\ba
&& {\nn=}\textrm{even}:\quad  {\cal S}[\lambda]  =  \diag \left( \mu, \frac{1}{\mu}\ ,\dots ,   \frac{1}{\mu}  \right) \ ,
\nonumber\\
&& {\nn=}\textrm{odd}:\quad\  {\cal S}[\lambda]  =  \diag \left( \frac{1}{\mu}  , \mu ,\dots , \frac{1}{\mu}  \right) \ ,
\quad \mu = \sqrt{1- \l  \ov 1+ \l }\ ,
\label{fram2}
\ea
in which we have taken into account an overall constant of $\sqrt{\l^{-2}-1}$ entering in the $\sigma$-model metric given by eq.~\eqref{eq:metriccoset}.
Classical integrability has been observed in $\sigma$-models on squashed
{\em geometric} cosets (i.e. the squashed three-sphere) in the literature
\cite{Orlando:2010yh,Kawaguchi:2011mz,Kawaguchi:2011pf,Orlando:2012hu}  but here we are talking about
squashing of a conformal coset preserving integrability - something quite different.

\no
Note also that the $\lambda \rightarrow 1$ limit of this procedure is, naively, bad since the tangent space metric degenerates.
However, the correct procedure, as we explained above around \eqn{glim}, is to perform a rescaling of the group element in such a way
 that the limit can be taken.  When this is done taking $\lambda \rightarrow 1$ drives towards the non-Abelian T-dual
 of the PCM on the geometric coset $S^\nn$  dualized with respect to $SO(\nn +1)$  \cite{Lozano:2011kb}.

\subsection{The $\nn=5$ case}

It remains to give the explicit form of the frame fields which are, in fact, profoundly complicated.
Here we present the results for $\nn=5$ which to our knowledge have not appeared in the literature even for the undeformed gauged WZW model.
Whilst the metric for the cases $\nn=3,4$ are available in the literature \cite{Bars:1991pt,Fradkin:1991ie,Bars:1991zt},
the frame fields in this basis in which the deformation
is a simple squashing had not been systematically given so for completeness we include them in the appendix.

\noindent
To simplify results we perform a coordinate transformation\footnote{With this transformation we will find that the metric
for the $\l=0$ coset CFT has no off-diagonal terms in the $du$ and $d\omega$ components.  Of course for $\l\neq 0$ the metric becomes intractably
off-diagonal.}
\ba
&&\om = \tan \phi  \ ,\quad  x = \sin \th_1\ , \quad y = \cos \th_1 \cos \th_2 \ ,
\nonumber\\
&& z = \cos \th_1 \cos \th_2 \cos \th_3 \ , \quad u =  \cos \th_1 \cos \th_2 \cos \th_3 \cos \th_4\ .
\ea
We also  introduce the functions
 \be
 \label{eq:shortcuts}
 \begin{aligned}
 {\cal A} &=  1- x^2 -y^2  \ , \quad
 {\cal B}  = y^2- z^2 \ , \quad
 {\cal C}  = z^2 - u^2 \ ,\quad
 {\cal D}  = 1- x^2 \ , \\
 {\cal E} &= u^2 y^4 + u^2 x^2 z^2 - y^4 z^2 \ ,  \quad
 {\cal F} = u^2 y^2 + z^4 \omega^2 \ .
 \end{aligned}
 \ee
 Then the frame fields are given by, i.e. \eqn{fram1} and \eqn{fram2}
 \be\label{eq:S5frames1}
e^\a_{(\l)}  = S^{\a\b}(\l) e^\b_{(0)} \ ,\qq S(\l)=  \diag \left( \frac{1}{\mu}  , \mu ,\frac{1}{\mu}  , \mu,  \frac{1}{\mu}  \right) \ ,
\quad \mu = \sqrt{1- \l  \ov 1+ \l }\ ,
 \ee
 where the frames for the undeformed metric corresponding to the coset CFT $SO(6)/SO(5)$ are
 \ba\label{eq:S5frames2}
&& e^1_{(0)} =- \frac{2 \sqrt{k}  x}{y z \sqrt{\A\B\C\D }\omega} \Big( \A\B u du - \B(u^2 x^2 -z^2 \omega^2 \D) {dx\ov x}
\nonumber\\
&& \phantom{xxxx} - (\E-z^4 \omega^2 \D) {dy\ov y}  - \A \F {dz\ov z} + \A\B\C \omega d\phi   \Big)\ ,
\nonumber\\
&& e^2_{(0)} = \frac{2  \sqrt{k}  }{y z \sqrt{\A\B\C  } \omega} \left( \A\B u du - \B x(u^2 + z^2 \omega^2  ) dx
- (\E+ x^2z^4 \omega^2) {dy\ov y} - \A\F {dz\ov z} + \A\B\C \omega d\phi   \right)\ ,
\nonumber\\
&& e^3_{(0)} = -\frac{2 \sqrt{k}   }{  z \sqrt{\B\C\D  } \omega} \left( \B u du +\frac{\B x u^2}{y^2}dx + \frac{\E}{y^3} dy
- \frac{1}{z}(\F- u^2\D  )dz +  \B\C \omega d\phi   \right)\ ,
\label{eq:frames}
\\
&& e^4_{(0)} = -\frac{2 \sqrt{k}   }{  y \sqrt{\C  } \omega} \left(   u du  + z \omega^2dz +  \C \omega d\phi   \right)\ ,
\nonumber\\
&& e^5_{(0)} = - \frac{2 \sqrt{k}   }{  z   \omega} \left(    du - u \omega d\phi   \right)\ .
\nonumber
\ea

Making use of \eqref{eq:DilDet} one sees that the dilaton receives only a constant shift
$\Phi_{\const} =   - \frac{1}{2} \ln \det {\bf P} =   \frac{1}{2} \ln  \frac{\sqrt{1- \l   } }{\sqrt{1+ \l }}$
 away from the gauged WZW dilaton. We find that
        \be
    e^{-2\Phi  + 2 \Phi_{\const} } =\frac{1024 \mathcal{A} \mathcal{C}^2 \omega ^4 \mathcal{B}}{(1+\omega^2)^4 z^2} \ .
        \ee
  As discussed before there is no NS $B$-field.

For this theory to make sense one should first check that the dilaton beta-function comes out as a constant in the deformed case.
One should be clear that even though the deformation looks, in tangent space, like a rather simple squashing,
is not obvious from the outset that this will work out;
indeed one finds rather quickly that the squashing wreaks havoc on the spin connection.
Nonetheless one finds that the dilaton equation gives
      \be\label{eq:Dilcomp}
      \begin{aligned}
  R+ 4 \Box \Phi -4 (\partial \Phi)^2 &= \frac{10}{k} \left( \frac{1+\lambda^2 }{1-\lambda^2 }\right) \ .
      \end{aligned}
      \ee
Hence, already a strong consistency check passed.  The Einstein equation yields
      \be
      \begin{aligned}
      e_a^\mu e_b^\nu \left( R_{\mu \nu} + 2 \nabla_\mu \nabla_\nu \Phi \right) &=\frac{4}{k(1-\lambda^2)} \diag (-1, 1,-1,1,-1)\ .
            \end{aligned}
      \ee
The fact that this is not satisfied indicates that we shall need to activate RR-fluxes in the attempt to find the full solution.
    The simple form of the stress tensor on the right hand side of the Einstein equations suggests that the
     RR-fields will be rather simple when written in terms of the natural frame fields \eqref{eq:frames}.

\section{A comment on analytic continuation}
\label{wick}

We are interested in finding a full supergravity embedding for which we will want to combine these results
with some corresponding non-compact coset. In essence this is achieved by choosing appropriate analytic continuations
of the corresponding compact geometry combined with sending the level $k \rightarrow -k$.  For gauged-WZW models this
procedure is quite well established  \cite{Bars:1991pt,Fradkin:1991ie}  and in the case of $\lambda$-deformations was used for
the examples of $SU(2)/U(1)$ and  $SO(4)/SO(3)$  in  \cite{Sfetsos:2014cea}.

    Suppose we start with some $SO(\nn+1)/SO(\nn)$ gauge-fixed element $g$ given by the Euler angles as in \eqref{gfixed}
    and make the continuation  $\theta \rightarrow i \theta$ of one of the angles.   The result will be that $g$ is now a complex
    matrix obeying $g^t = g^{-1}$.  But what we wanted was to end up with a real element of e.g.  $SO(\nn-1,2)$ gauged fixed
    under the action of $SO(\nn-1,1)$. Suppose we find a matrix $\rho$ such that $\tilde{g}= \rho g \rho^{-1}$ is real,
    then $\tilde{g}$ will preserve the metric given by $\eta = \rho^T \rho$ and will  be an appropriate gauge-fixed element
    of a non-compact group $G$ quotiented by the gauge action of a non-compact subgroup $H$.  The original generators are
    transformed as $T\rightarrow \rho T \rho^{-1}$ leading to a certain number of non-compact generators inside $G$ and $H$.
    The  signature of the corresponding $\s$-model spacetime is deduced by seeing how many time-like directions lie in
    the subspace in which the subgroup $H$ acts.   To end up with a theory with only one time-like direction this procedure
    is supplemented by sending the level $k \rightarrow -k$.  As geometric cosets the resulting $SO(\nn-1,2)/SO(\nn-1,1)$ are of Anti-de-Sitter type. De-Sitter cosets can be realised in a similar fashion; rotating angles and performing an action $\rho$ to give $SO(\nn,1)/SO(\nn-1,1)$; the difference in this case is that one does {\em not} need to switch the sign of the level to end up with a single time-like direction.

There can be many different analytic continuations that all give rise to geometries of the same signature which differ in
how non-compact generators are assigned.  Indeed to end up in Minkowski signature starting from an $\nn$-dimensional Euclidean theory,
we can find $\nn$ different analytic continuations depending on which of the $\nn$ directions in tangent space we wish to make time-like.
The results for  $\nn=5$, the case of interest in the current note, are summarised in   table 1.

\begin{table}[h!]
\label{table:Wick}\centering
  \begin{tabular}{cccc}
    \hline
   G/H & Angles Rotated & Signature & Type  \\
    \hline
$SO(5,1)/SO(5)$ & $\phi $ & $\{+,+,+,+,+\}$  & EAdS  \\
$SO(4,2)/SO(4,1)$ & $\theta_4$ & $\{+,+,+,+,-\}$ &  AdS    \\
$SO(4,2)/SO(4,1)$ & $\theta_4,\theta_3, \phi$ &  $\{+,+,+,-,+\}$ &  AdS  \\
$SO(4,2)/SO(4,1)$ & $\theta_3, \theta_2, \phi$ &  $\{+,+,-,+,+\}$   &  AdS\\
$SO(4,2)/SO(4,1)$ & $ \theta_2, \theta_1,  \phi$ &  $\{+,-,+,+,+\}$   &  AdS \\
$SO(4,2)/SO(4,1)$ & $\theta_1,  \phi$ & $\{-,+,+,+,+\}$  &  AdS    \\
$SO(5,1)/SO(4,1)$ & $\theta_1$ & $\{-,+,+,+,+\}$  &   dS    \\
$SO(5,1)/SO(4,1)$ & $\theta_2, \theta_1$ & $\{+,-,+,+,+\}$  &   dS    \\
$SO(5,1)/SO(4,1)$ & $\theta_3, \theta_2$ & $\{+,+,-,+,+\}$  &   dS    \\
$SO(5,1)/SO(4,1)$ & $\theta_4, \theta_3$ & $\{+,+,+,-,+\}$  &   dS    \\
$SO(5,1)/SO(4,1)$ & $\theta_4, \phi$ & $\{+,+,+,+,-\}$  &   dS    \\
    \hline
  \end{tabular}
\caption{Analytic continuations for $n=5$; the  signature  column  indicates which of the tangent
space directions becomes time-like in the basis of frames in eq.~\eqref{eq:frames}.  For completeness we include the rotations that would correspond to Euclidean AdS, AdS and dS  type geometric cosets.}
\end{table}

 %

\section{The supergravity embedding for $AdS_5\times S^5$}
\label{sugra}

We are interested in understanding the circumstances in which we are able to find supergravity
embeddings obtained by combining the deformation of a compact coset  CFT with that of a non-compact coset CFT.
For $\nn=3$ (also for $\nn=2$ in which case the analytic continuation is quite simple)
this was done in \cite{Sfetsos:2014cea} giving the $\lambda$-deformation to the coset CFT of
$\frac{SO(2,2)}{SO(2,1)}\times \frac{SO(4)}{SO(3)} \times U(1)^4$.  Taking the deformation parameter
$\lambda \rightarrow 1$ (with the previously discussed rescaling the group element) one recovers the
non-Abelian T-dual of the PCM for strings on $AdS_3\times S^3$ supported by RR flux.
Here we will consider $\nn=5$.

An important question is whether the bosonic geometry can be supported by {\em real} RR-fields;
this would certainly make the interpretation simpler than finding oneself in a Type-$II^{\star}$  theory.
To address this we need to use some intuition for the structure of the RR-fields.   We are not going to
attempt a first principle derivation but to bootstrap a solution from knowledge of the bosonic sector
and compatibility with the end points of the $\lambda$-deformation.  For $\lambda=0$ the RR-fields should
vanish and for the $\lambda=1$ limit they should match those of non-Abelian T-duality determined in \cite{Sfetsos:2010uq,Lozano:2011kb}.
The conservative ansatz, which actually works, is to assume that they have the same structure as those obtained via
non-Abelian T-duality but are multiplied by a suitable overall function of the deformation parameter.

 In T-duality (Abelian and non-Abelian alike) the transformation for RR-fields arises due to the
 fact that left and right movers on the world-sheet couple to different frame fields after T-duality and this induces a transformation on spinors \cite{Hassan:1999bv,Sfetsos:2010uq,Lozano:2011kb,Kelekci:2014ima}.
 This was also the case in the $\lambda$-deformation; both the left- and right-moving gauge fields defined a
 set of frame fields related by a Lorentz transformation given by eq.~\eqref{eq:Lambda2} in general. For the  case of $SO(6)/SO(5)$, due to the simple form of the ${\bf P}$ matrix, this Lorentz transformation reduces to
 \be\label{eq:Lambda3}
 \Lambda = \diag\left(-1, 1, -1 , 1,-1 \right) \ .
 \ee
This is the same relation between left and right moving frame fields as would be obtained by performing
three Abelian T-dualities giving reflections in the $e^1, e^3$ and $e^5$ directions.  Lets temporarily
pretend that we are just doing T-dualities.  In general performing a T-duality in a time-like direction produces imaginary RR-fluxes
 \cite{Hull:1998vg}. In order to avoid such a situation we choose an analytic continuation in which $e^1, e^3$ and $e^5$
 remain space like.  From table~1, we see two possibilities either rotate angles  $\{\theta_4, \theta_3, \phi\}$ or  $\{\theta_2, \theta_1, \phi\}$.
 For specificity and also because this is the rotation can be used in lower dimensions as in  \cite{Sfetsos:2014cea}  we choose the later.

The continuation on  $\{ \theta_1, \theta_2,  \phi$ \}   is achieved by sending $\{x , y , z, u, \omega\}
 \rightarrow \{  i \tilde x, \tilde y,  \tilde z, \tilde u, i \tilde \omega\}$ in the Cartesian coordinate system and assigning ranges:
\begin{equation}
\tilde{y}^2- \tilde{x}^2 > 1 \ , \quad \tilde{z}^2 - \tilde{u}^2> 0 \ , \quad \tilde{u}^2 - \tilde{y}^2 <0 \ , \quad -1< \tilde\omega <1  \ .
\end{equation}
In addition, $\tilde\Phi$ must also be rendered real by subtracting of an $i \pi/2$ that comes from a negative sign
inside a logarithm and arrive at real frame fields by defining the  time-like direction   $\tilde{e}^0 = - i \tilde e^2$.
This process defines frame fields  $\{\tilde e^0 , \tilde e^1 ,\tilde e^3, \tilde e^4 , \tilde e^5 \}$ for the non-compact geometry given explicitly by,
 \be\label{eq:Ads5frames1}
\tilde{e}^\a_{(\l)}  = S^{\a\b}(\l) \tilde{e}^\b_{(0)} \ ,\qquad \a = 1 \dots 5 \ ,
 \ee
 where the deformation matrix $S(\lambda)$ was defined in eq.~\eqref{eq:S5frames1} and where the undeformed frames are,
 \def\At{\tilde{{\cal A}}}
  \def\Bt{\tilde{{\cal B}}}
   \def\Ct{\tilde{{\cal C}}}
    \def\Dt{\tilde{{\cal D}}}
     \def\Et{\tilde{{\cal E}}}
      \def\Ft{\tilde{{\cal F}}}
      \def\xt{\tilde{x}}
      \def\yt{ \tilde{y}}
      \def\zt{\tilde{z}}
      \def\ut{\tilde{u}}
      \def\wt{\tilde{\omega}}
            \def\pt{\tilde{\phi}}
 \begin{equation}
 \label{eq:AdS5frames2}
 \begin{aligned}
 & \tilde e^1_{(0)} =- \frac{2 \sqrt{k}  \xt}{\yt \zt \sqrt{\At\Bt\Ct\Dt }\wt} \Big( \At \Bt \ut d\ut - \Bt (\ut^2 \xt^2 -\zt^2 \wt^2 \Dt) {d\xt\ov \xt} + (\Et + \zt^4 \wt^2 \Dt) {d\yt \ov \yt}  - \At \Ft {d\zt \ov \zt} - \At\Bt\Ct \wt d\pt   \Big)\ ,
 \\
  &\tilde e^0_{(0)}\equiv  - i  \tilde e^2_{(0)}        = \frac{2  \sqrt{k}  }{\yt \zt \sqrt{\At\Bt\Ct  } \omega} \left( -\At\Bt \ut d\ut + \Bt \xt(\ut^2 - \zt^2 \wt^2  ) d\xt
- (\Et+ \xt^2 \zt^4 \wt^2) {d\yt \ov \yt}  + \At\Ft {d\zt\ov \zt} + \At\Bt\Ct \wt d\pt   \right),
 \\
 &  \tilde  e^3_{(0)} = -\frac{2 \sqrt{k}   }{  \zt \sqrt{\Bt\Ct\Dt  } \wt } \left( \Bt \ut d\ut - \frac{\Bt \xt \ut^2}{\yt^2}d\xt + \frac{\Et}{\yt^3} d\yt
- \frac{1}{\zt}(\Ft- \ut^2\Dt  )d\zt -  \Bt\Ct \wt d\pt   \right)\ ,
\\
 & \tilde e^4_{(0)} = -\frac{2 \sqrt{k}   }{  \yt \sqrt{\Ct  } \wt} \left(   \ut d\ut  -\zt  \wt^2d\zt -  \Ct \wt d\pt   \right)\ ,
 \\
   & \tilde e^5_{(0)} = - \frac{2 \sqrt{k}   }{  \zt   \wt } \left(    d\ut +\ut \wt d\pt  \right)\ .
  \end{aligned}
  \end{equation}
  Here we have defined the functions  
 \be
 \label{eq:shortcuts2}
 \begin{aligned}
\tilde {\cal A} &= \yt^2 - \xt^2 -1    \ , \quad
\tilde {\cal B}  = \yt^2- \zt^2 \ , \quad
\tilde {\cal C}  = \zt^2 - \ut^2 \ ,\quad
\tilde {\cal D}  = 1 + \xt ^2 \ , \\
\tilde  {\cal E} &= \ut^2 \yt^4 - \ut^2 \xt^2 \zt^2 - \yt^4 \zt^2 \ ,  \quad
\tilde  {\cal F} = \ut^2 \yt^2 -  \zt^4 \wt^2 \ ,\quad \wt = \tanh \tilde\phi \  .
 \end{aligned}
 \ee
  
  One finds  a dilaton equation for this geometry
\be
 \tilde R+ 4 \tilde\Box \tilde\Phi -4 (\partial \tilde\Phi)^2   = - \frac{10}{k} \left( \frac{1+\lambda^2 }{1-\lambda^2 }\right)\ ,
  \ee
 which provides an exact cancelation with the contribution coming from the compact space in eq.~\eqref{eq:Dilcomp}.

 We thus define the full type-IIB supergravity embedding by combining the compact part defined by the frames in eq.~\eqref{eq:frames} with the above non-compact part in eq.  \eqref{eq:AdS5frames2}  reordered such that the time like direction comes first:
\begin{equation}
\frak{e}^I = \{- i \tilde e^2 , \tilde e^1 ,\tilde e^3, \tilde e^4 , \tilde e^5 , e^1 ,e^2 ,e^3 ,e^4 , e^5\}  \ , \quad I = 0,1, \dots, 9 \ .
\end{equation}
The full dilaton is
\begin{equation}
\begin{aligned}
\Phi =& -\frac{1}{2} \ln \left(-\frac{  \omega ^4 \left(u^2-z^2\right)^2
   \left(x^2+y^2-1\right) (y^2-z^2) }{\left(\omega ^2+1\right)^4 z^2}\right)  \\
 &  -\frac{1}{2} \ln \left(\frac{\tilde{\omega}^4
   \left(\tilde{u}^2-\tilde{z}^2\right)^2 \left(-\tilde{x}^2+\tilde{y}^2-1\right)
   (\tilde{y}^2-\tilde{z}^2) }{\left(1-\tilde{\omega}^2\right)^4
   \tilde{z}^2}\right) + \phi_0 \ .
   \end{aligned}
\end{equation}
There is no NS two-form and the dilaton supergravity equation is solved.
 The RR-fields follow from the ansatz advocated in  \cite{Sfetsos:2014cea}; the frame rotation \eqref{eq:Lambda3}
 together with the analogous rotation in the non-compact part of the theory give rise to an action on spinors
 through $\Omega \Gamma^I \Omega^{-1} = \Lambda^I{}_J  \Gamma^J$  with
 \be
\Omega = \Gamma^{124}\Gamma^{579}  \ .
\ee
We then construct
\be
\widehat{\slashed{F}}= f(\lambda) e^{\Phi}\slashed{F}_0\cdot  \Omega   \ ,
\ee
where the slashed notation indicates contraction with $\Gamma$ matrices and where
\be
 (F_5)_0 = \frak{e}^{01234} -  \frak{e}^{56789}\ ,
\ee
is inherited from the type-IIB $AdS_5\times S^5$ geometry and $f(\lambda)$ is a function of the deformation
 parameter that is fixed from the Einstein equations.   The result is that we find the following five-form flux
\be
\widehat{F}_5 = e^{-\Phi} \frac{4 \sqrt{\lambda}}{ \sqrt{k(1-\l^2) }} \left(  \frak{e}^{03579} -  \frak{e}^{12468} \right) \ .
\ee
This is self-dual and one can check it solves its Bianchi identity.  With this one also finds that the
Einstein equations are completely solved.

\section{Conclusions}
Let us summarize what we have learnt.   The first lesson is that for cases of $SO(\nn +1 )/SO(\nn)$,
the $\lambda$-deformation has a very natural action; it simply corresponds to squashing certain tangent
space directions in the metric.   An indication that this makes sense is that at a quantum level this
gives a constant shift to the dilaton beta-function.    The second main point is that the $\lambda$-deformed
 $SO(\nn +1 )/SO(\nn)$ gauged-WZW can be coupled via a non-trivial Ramond--Ramond sector to a similarly
 deformed  $SO(\nn -1,2 )/SO(\nn-2, 1 )$ theory in such a way that the supergravity fields solve all the one-loop beta-function equations.

In general one should be able to derive these results from a direct application to the superstring.
There is a simple form due to Hollowood, Miramontes and Schmidtt \cite{Hollowood:2014qma} for the action of the
$\lambda$-deformation in terms of a deformed $G/G$ gauged WZW for $F= PSU(2,2|4)$,
  \be
  \label{eq:GSaction}
  \begin{aligned}
  S_{\rm def}[F,A] = S_{gWZW}[F,A] - \frac{k}{\pi} \int d^2 x {\mathrm{STr}}\left[ A_+ \left( \Omega - \mathbb{1} \right) A_- \right] \ ,  \\
  \Omega =  \mathbb{P}_0 + \lambda^{-1} \mathbb{P}_1+\lambda^{-2}  \mathbb{P}_2 + \lambda \mathbb{P}_3 \ ,
  \end{aligned}
  \ee
    in which $ \mathbb{P}_i$ are the projectors onto the eigenspaces of the usual $\mathbb{Z}_4$
    automorphism. The powers of $\l$ entering in this $\Omega$ are tuned precisely such that this theory admits a Lax formulation.
      In the present paper we have considered a bosonic truncation of this theory that
    shares exactly the same NS-sector.   By completing this bosonic sector with RR-fields to give the
    supergravity embedding, we have given some strong supporting evidence that the
    $\lambda$-deformation of superstrings based on the $PSU(2,2|4)$ supercoset is a marginal deformation.

There are,  however, a few caveats that remain: the first is that whilst the
RR-fields we have used are natural, and are suggested by an ansatz known from considerations of non-Abelian T-duality,
 one can not be {\em absolutely} certain that they correspond to the RR-sector obtained by performing the
  $\lambda$-deformation directly in the $PSU(2,2|4)$ super-coset. ({\bf See note added}.)
  Our solution of supergravity is bootstrapped from a precise knowledge of the bosonic sector and
  unfortunately it is exceedingly difficult to extract the RR-fields directly from $\sigma$-models
  \eqref{eq:GSaction} to compare with.  It may be that a direct calculation of the
   renormalisation of eq.~\eqref{eq:GSaction} is required.
     A second issue concerns the target space interpretation.  To obtain the RR-sector with real
     fluxes one has to pick an appropriate Wick rotation and moreover we have contented ourselves
     to work on a particular patch.  It would be interesting to explore how global extensions of
     the geometries we consider can be supported by RR-fluxes and whether they remain real.
   It is known from the work and the examples worked out in \cite{Sfetsos:2014cea} that this is a non-trivial issue. 

  There are a number of open avenues that we believe deserve exploration:
  \begin{itemize}
  \item Can one develop a more general theory of squashing conformal cosets and their geometrical properties? In that respect
 we recall the work in \cite{Petropoulos:2006py} in which hierarchies of non-Abelian coset CFTs of orthogonal groups were constructed
via asymptotic limits. It will be interested to investigate if these structures survive the $\l$-deformations.

  \item Can the $\lambda$-deformation be applied to other scenario's where integrability is expected e.g. $AdS_4 \times CP^3$?
  \item What can one say in general about multi-parameter or anisotropic $\lambda$-deformations in which  the deformation parameter is replaced with a matrix $\l \mathbb{1}_{AB} \rightarrow \l_{AB}$?
  \item    It has been shown that $\l$-deformations of WZW models have a an underlying Yangian symmetry \cite{Itsios:2014vfa}.
  Would it be possible to establish in a similar fashion the expected quantum group symmetries for $\l$-deformations of coset CFT models   and in particular for
  $AdS_5 \times S^5$?

  \item How can we make more precise the linkage between the $\eta$-deformations of \cite{Delduc:2013qra} and these $\lambda$-deformations ? {\bf See note added}.
    \end{itemize}
  Of course the grandest, and most tantalizing question of all:
  {\em What do both $\eta$ and $\lambda$ deformations imply  for ${\cal N}=4$ SYM?}

\section*{Acknowledgements}

  We thank  B. Hoare, A. Sevrin and J. Vanhoof  for helpful discussions.
The research of K. Sfetsos is implemented
under the \textsl{ARISTEIA} action (D.654 of GGET) of the \textsl{operational
programme education and lifelong learning} and is co-funded by the
European Social Fund (ESF) and National Resources (2007-2013).
 The work of DCT was supported in part by FWO-Vlaanderen through project G020714N
and postdoctoral mandate 12D1215N, by the Belgian Federal Science Policy Office through the Interuniversity Attraction Pole P7/37,
and by the Vrije Universiteit Brussel through the Strategic Research Program ``High-Energy Physics''.
\begin{appendix}

 \section*{Note Added}
 
 After this article appeared   on the arXiv but before going to press we received two preprints \cite{Vicedo:2015pna,Hoare:2015gda} that address some of the points raised in the conclusion.  
 
In \cite{Vicedo:2015pna} it is demonstrated that  Yang-Baxter deformations  on the real branch for symmetric cosets are Poisson-Lie T-dual to the sorts of $\lambda$-deformations considered in this letter.  The $\eta$-deformation is of Yang-Baxter type but on the complex  branch (here complex and real essentially refer to a sign choice made in the modified Yang-Baxter equation determining a choice of R-matrix). However, it is expected that a combination of Poisson-Lie T-duality and analytic continuation directly relates the $\eta$-deformation to the $\lambda$-deformation; this was shown explicitly for the case based on $SU(2)/U(1)$ in \cite{Hoare:2015gda}.

A second point concerns the relation to the geometry found within and that corresponding to the $\sigma$-model of \cite{Hollowood:2014qma}.  The prescription given for the dilaton in  \cite{Hollowood:2014qma} does not immediately correspond to that coming from eq.~\eqref{eq:DilDet}.   It was shown explicitly in  \cite{Hoare:2015gda} for the $AdS_3\times S^3$ $\lambda$-deformation  that the fermionic contributions from the supercoset produce an additional contribution to the dilaton over that found in  \cite{Sfetsos:2014cea}.  The expression given for the dilaton in \cite{Hoare:2015gda} is significantly more complicated than that of \cite{Sfetsos:2014cea} but nonetheless the dilaton equation of motion and the trace Einstein equations remain solved; the corresponding form of the RR fluxes is not yet known but it seems that these too will receive complicated corrections for the fermionic terms in the supercoset.   The situation appears to be rather comparable to fermionic T-duality which preserves the NS sector but modifies the dilaton and RR sector of a background; presumably by performing the $\lambda$-deformation in the fermionic directions one would receive a correction to the RR sector.  We hope to return to these issues.

 We thank the authors of \cite{Vicedo:2015pna,Hoare:2015gda} for  email correspondence on these points.
 
\section{Appendix}
Here we present the frame fields for the $\lambda$-deformed $SO(\nn+1)/SO(\nn)$ for $\nn = 2,3,4$.
 We use the same Cartesian coordinates and definitions referred to in the main text in (\ref{eq:shortcuts}) and in addition
 define $\lambda_\pm = \sqrt{k(1\pm \lambda)/(1\mp \lambda)}$ and $\omega_+^2 = 1+ \omega^2$.   A potentially useful observation is that these geometries are nested;
 for instance for $\lambda=0$ setting $u=0$, $du=0$ and sending $\phi \rightarrow \phi+ \frac{\pi}{2} $
 one has $ds^2_{\nn=5} \rightarrow ds^2_{\nn=4}$ perhaps hinting at a   connection with \cite{Grigoriev:2007bu}.

\subsection{The $\nn=2$ case}
The deformed frame fields read 
\be
  \begin{aligned}
  e^1 &= - \frac{2 \lambda_- dx}{\omega } + \frac{2 x  \lambda_- d\omega }{ \omega_+^2},\\
  e^2 &=-   \frac{2 \lambda_+ x dx}{ \sqrt{\mathcal{D}}\omega } - \frac{2  \lambda_+\sqrt{\mathcal{D}}  d\omega}{ \omega_+^2}.
   \end{aligned}
 \ee
The non-constant part of the dilaton is
    \be
    e^{-2\Phi} = \frac{2 \omega ^2}{\omega_+^2}.
    \ee
The spinorial counterpart of the Lorentz rotation between left and right moving frames is
   \be
      \Omega =  \G_2.
      \ee
  The dilaton beta function equation is
 \be
\beta^\Phi = R + 4 \nabla^2 \Phi - 4 (\partial \Phi)^2  = \frac{1}{k} \frac{1+\l^2}{1-\l^2}.
\ee

\subsection{The $\nn=3$ case}
For $SO(4)/SO(3)$ the deformed frame fields read 
\be
  \begin{aligned}
e^1 &= - \frac{2 \lambda_+ \left(\mathcal{D} \omega ^2  \omega_+^2 dx+x \left(y  \omega_+^2 dy + \mathcal{A} \omega  d\omega \right)\right)}{\sqrt{\mathcal{A}\mathcal{D}} \omega   \omega_+^2}\ ,\\
 e^2 &= + \frac{2 \lambda_- \left(-x
   \omega ^2  \omega_+^2 dx+y  \omega_+^2 dy+\mathcal{A} \omega  d\omega\right)}{\sqrt{\mathcal{A}} \omega   \omega_+^2}\ ,\\
  e^3 &= +  \frac{2 \l_+
   \left( \omega_+^2 dy-y \omega  d\omega\right)}{\sqrt{\mathcal{D}} \omega   \omega_+^2}\ .
   \end{aligned}
 \ee
The dilaton, spinorial Lorentz rotation and the dilaton beta function yield
        \be
    e^{-2\Phi} =\frac{8 \mathcal{A} \omega ^2}{\omega_+^4},
        \ee
        \be
      \Omega =  \G_1\Gamma_3,
      \ee

\be
\beta^\Phi =     \frac{3}{k} \frac{1+\l^2}{1-\l^2}.
\ee
\subsection{The $\nn=4$ case}
The four frame field are given by
   \be
  \begin{aligned}
 e^1&= - \frac{2 \lambda_-\left( \mathcal{D} y  \omega_+^2 \mathcal{B} dx+x \left( \omega_+^2  \left(\mathcal{D} z^2+y^4 \omega
   ^2\right)dy - \mathcal{A} y \left( \omega_+^2 z dz+\omega  \mathcal{B} d\omega \right)\right)\right)}{\sqrt{\mathcal{A}\mathcal{B}\mathcal{D}} 
   \omega   \omega_+^2 y^2}\ ,\\
  e^2&=  + \frac{2 \lambda_+\left( x y
    \omega_+^2 \mathcal{B} dx+\omega_+^2  \left(x^2 z^2-y^4 \omega ^2\right)dy+\mathcal{A} y \left( \omega_+^2 z dz+\omega  \mathcal{B} d\omega \right)\right)}{\sqrt{\mathcal{A}\mathcal{B}}
   \omega   \omega_+^2 y^2}\ ,\\
 e^3&=+   \frac{2 \lambda_-\left(y \omega ^2  \omega_+^2 dy+ \omega_+^2 z dz+\omega
   \mathcal{B} d\omega \right)}{\sqrt{\mathcal{B}\mathcal{D}} \omega   \omega_+^2 }\ ,\\
 e^4&= -  \frac{2 \l_+
   \left( \omega_+^2 dz-z \omega  d\omega\right)}{ y \omega   \omega_+^2}  \ .
   \end{aligned}
 \ee
While the dilaton, spinoral Lorentz rotation and the dilaton beta function respectively read
     \be
    e^{-2\Phi} =\frac{64 \mathcal{A} \omega ^4 \mathcal{B}}{\omega_+^6},
        \ee
           \be
      \Omega =  \G_2\Gamma_4,
      \ee
\be
\beta^\Phi =     \frac{6}{k} \frac{1+\l^2}{1-\l^2}.
\ee

\end{appendix}

 \end{document}